\documentclass[traditabstract]{aa} 
\usepackage{graphicx}
\usepackage{txfonts}
\usepackage{natbib}
\usepackage{graphicx}
\usepackage{gensymb}
\usepackage{multirow}

\bibpunct{(}{)}{;}{a}{}{,} 

\bibpunct{(}{)}{;}{a}{}{,}
\def\aap{A\&A}                
\def\ao{Applied Optics}       
\def\apjl{ApJ}                
\def\apjs{ApJS}              
\newcommand{\chem}{\everymath={\fam0 }\fam0 }  
\newcommand{\iso}[2]{{}^{#1}_{#2}}

\title{Mass entrainment and turbulence-driven acceleration\\ of ultra-high energy cosmic rays in Centaurus\,A}

\author{Sarka Wykes \inst{1,2}$^{,}$\thanks{corresponding author: sarka@astro.ru.nl}
\and Judith H. Croston \inst{3} 
\and Martin J. Hardcastle \inst{4} 
\and Jean A. Eilek \inst{5,6}
\and Peter L. Biermann \inst{7,8,9,10}
\and \\Abraham Achterberg \inst{1} 
\and Justin D. Bray \inst{11,12}
\and Alex Lazarian \inst{13}
\and \\Marijke Haverkorn \inst{1,14}
\and Ray J. Protheroe \inst{11}
\and Omer Bromberg \inst{15} 
}

\authorrunning{Sarka Wykes et al.}
\titlerunning{Entrainment and UHECR acceleration in Cen\,A}

\institute{Department of Astrophysics/IMAPP, Radboud University Nijmegen, P.O. Box 9010, 6500 GL Nijmegen, The Netherlands
\and Astronomical Institute `Anton Pannekoek', University of Amsterdam, P.O. Box 94249, 1090 GE Amsterdam, The Netherlands
\and School of Physics and Astronomy, University of Southampton, University Road, Southampton, Hampshire SO17 1BJ, UK
\and School of Physics, Astronomy and Mathematics, University of Hertfordshire, College Lane, Hatfield, Hertfordshire AL10 9AB, UK
\and National Radio Astronomy Observatory, Socorro NM 87801, USA
\and Physics Department, New Mexico Tech, Socorro NM 87801, USA
\and Max-Planck-Institut f\"ur Radioastronomie, Auf dem H\"ugel 69, 53121 Bonn, Germany
\and Institute for Nuclear Physics, Karlsruhe Institute of Technology, P.O. Box 3640, 76021 Karlsruhe, Germany
\and Department of Physics and Astronomy, University of Alabama, Tuscaloosa, AL 35487, USA
\and Department of Physics, University of Alabama at Huntsville, Huntsville, AL 35899, USA
\and School of Chemistry \& Physics, University of Adelaide, SA 5005, Australia
\and CSIRO Australia Telescope National Facility, P.O. Box 76, Epping NSW 1710, Australia
\and Department of Astronomy, University of Wisconsin, 475 North Charter Street, Madison, WI 53706, USA
\and Sterrewacht Leiden, Leiden University, P.O. Box 9513, 2300 RA Leiden, The Netherlands
\and Racah Institute of Physics, The Hebrew University, Jerusalem 91904, Israel
}

\date{Received 2 April 2013 / Accepted 1 August 2013}

\begin{document}

\abstract {Observations of the FR\,I radio galaxy Centaurus\,A in radio, X-ray and gamma-ray bands provide evidence for lepton acceleration up to several TeV and clues about hadron acceleration to tens of EeV. Synthesising the available observational constraints on the physical conditions and particle content in the jets, inner lobes and giant lobes of Centaurus\,A, we aim to evaluate its feasibility as an ultra-high-energy cosmic-ray source. We apply several methods of determining jet power and affirm the consistency of various power estimates of $\sim1\times10^{43}$\,erg\,s$^{-1}$. Employing scaling relations based on previous results for 3C\,31, we estimate particle number densities in the jets, encompassing available radio through X-ray observations. Our model is compatible with the jets ingesting $\sim3\times10^{21}$\,g\,s$^{-1}$ of matter via external entrainment from hot gas and $\sim7\times10^{22}$\,g\,s$^{-1}$ via internal entrainment from jet-contained stars. This leads to an imbalance between the internal lobe pressure available from radiating particles and magnetic field, and our derived external pressure. Based on knowledge of the external environments of other FR\,I sources, we estimate the thermal pressure in the giant lobes as $1.5\times10^{-12}$\,dyn\,cm$^{-2}$, from which we deduce a lower limit to the temperature of $\sim1.6\times10^8$\,K. Using dynamical and buoyancy arguments, we infer $\sim440-645$\,Myr and $\sim560$\,Myr as the sound-crossing and buoyancy ages of the giant lobes respectively, inconsistent with their spectral ages. We re-investigate the feasibility of particle acceleration via stochastic processes in the lobes, placing new constraints on the energetics and on turbulent input to the lobes. The same `very hot' temperatures that allow self-consistency between the entrainment calculations and the missing pressure also allow stochastic UHECR acceleration models to work.}

\keywords{acceleration of particles -- cosmic rays -- galaxies: active -- galaxies: individual: Centaurus\,A (NGC\,5128) -- galaxies: jets -- turbulence}

\maketitle

\section{Introduction} \label{sect:introduction}

Relativistic jets and giant lobes of radio galaxies are potential sources of ultra-high energy cosmic rays (UHECRs) and very-high energy (VHE) neutrinos (e.g. \citealp{CAV78, BIE87, STE91, MAN95, BEN08, HAR09, KAC09, HAR10b, PEE12}). The synchrotron and inverse-Compton emission from these structures is seen in all wavebands from low-frequency radio to TeV gamma-ray. Knowledge of the physical conditions in jets and giant lobes is vital for understanding high-energy particle acceleration in full. Such conditions include the mean magnetic field strength and some estimate of its spatial variation, and the plasma densities both thermal and non-thermal.

\begin{figure}[!ht]
\begin{center}
\includegraphics[width=0.5\textwidth]{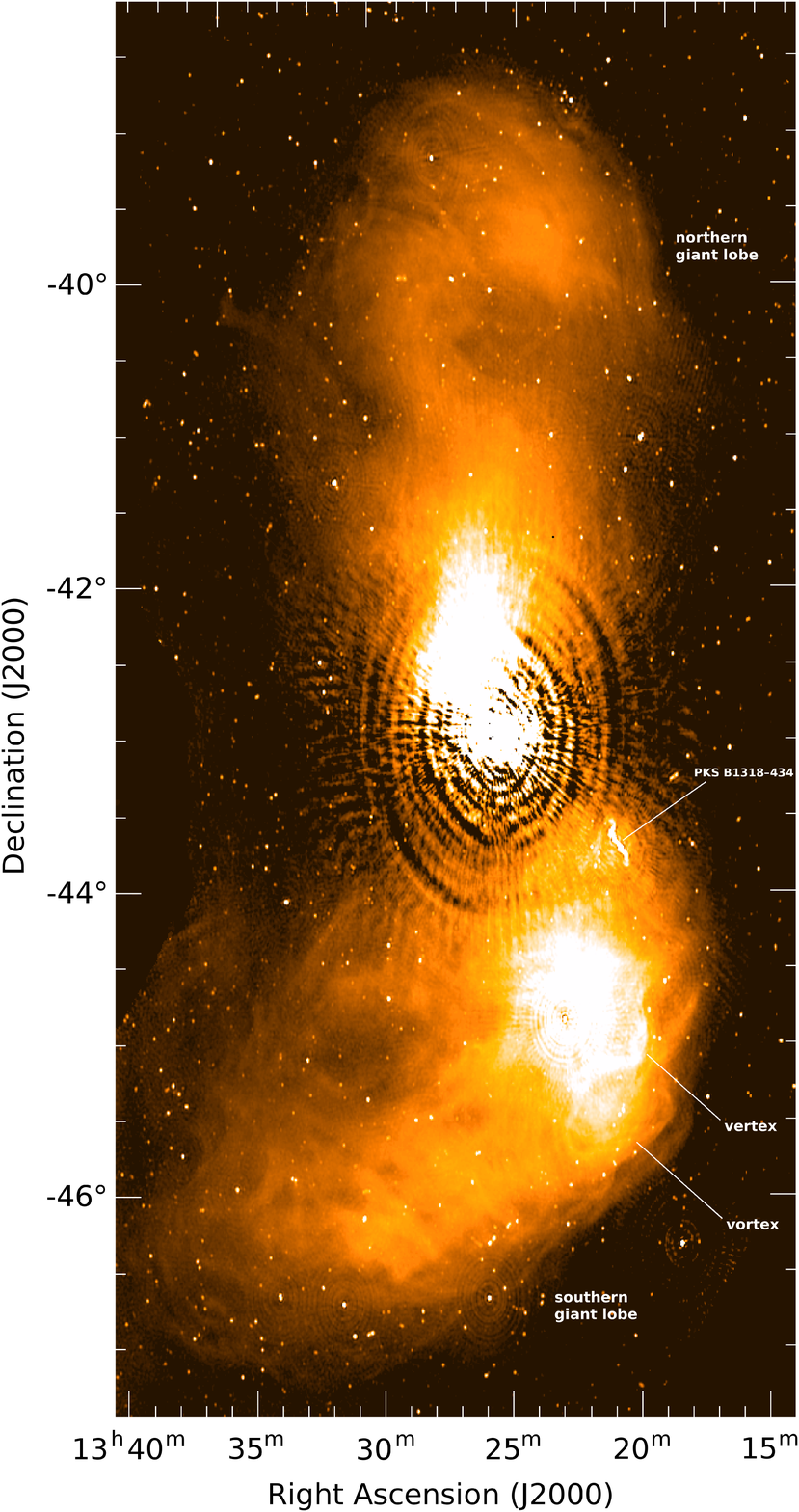}
\caption{Combined Australia Telescope Compact Array (ATCA) and Parkes 1.4\,GHz radio continuum image at $60''\times40''$ angular resolution of the large-scale structure of Centaurus\,A (adapted from \citealp{FEA11}), showing the giant lobes and the {\it vertex} and {\it vortex} filaments. The jets, inner lobes and the northern middle lobe are located in the saturated region centred on the nucleus. The elongated feature in the north-west part of the southern giant lobe, aligned approximately with the inner lobes, is the background FR\,I radio galaxy PKS\,B1318--434 of the Centaurus cluster at $\sim$ 45\,Mpc.}
\label{fig:cena}
\end{center}
\end{figure}

Results from the Pierre Auger Observatory (PAO) indicate that the UHECR composition changes as a function of energy \citep{PAO11a, PAO13} and that a number of the detected UHECRs could originate in the radio galaxy Centaurus\,A \citep{ABR10b}. Centaurus\,A (Cen\,A) is the nearest (3.8$\,\pm\,$0.1\,Mpc; \citealp{HARR10}) radio galaxy, a Fanaroff-Riley class\,I (FR\,I) object \citep{FAN74}, associated with the massive elliptical galaxy NGC\,5128, at the positional and dynamical center of the Centaurus group. Due to its brightness and proximity, Cen\,A is an excellent laboratory for detailed studies of particle acceleration, production of UHECRs, and the evolution of low-power radio galaxies in general. Several authors \citep{MOS09, HAR09, SUL09, GOP10, PEE12} have alluded to the possibility that the production of UHECRs occurs at Cen\,A's intermediate to large scales. \cite{HAR09}, \cite{SUL09} and more recently \cite{SUL11a} have considered stochastic acceleration by magnetohydrodynamic (MHD) turbulence as the driver for this in the large-scale lobes. The southern giant lobe is particularly interesting: it seems to be detaching or detached from the rest of the source and it features two prominent filaments, named the {\it vertex} and {\it vortex} (Fig.\,\ref{fig:cena}), which are the brightest (in flux density) filamentary structures known in any radio galaxy. Their origin has been suggested by \cite{FEA11} to be due to enhanced core/jet activity of the parent AGN or the passage of the dwarf irregular galaxy KK\,196, a Centaurus group member at $3.98\pm0.29$\,Mpc \citep{KAR07}, through the lobe.  

The jets in various FR\,I radio galaxies have been successfully described as turbulent, entraining, decelerating flows (e.g. 3C\,31, \citealp{LAI02a, PER07, WAN09}; B2\,0326+39 and B2\,1553+24, \citealp{CAN04}; NGC\,315, \citealp{CAN05}; 3C\,296, \citealp{LAI06}) with the entrainment process strongly affecting the evolution of the source. The approaching (i.e. northern) jet in Cen\,A has been traced out to a projected length of $\sim5$\,kpc in radio and 4.5\,kpc in X-rays, and from the changing properties of the X-ray emission at about 3.7\,kpc from the nucleus, \cite{HAR06} have alleged that the approaching jet enters the northern inner lobe at that point. Based on deep {\it Chandra} observations, \cite{HAR07} have claimed that the receding jet extends out to $\sim2$\,kpc in projection in X rays, and it also shows up on a similar scale in radio (\citealp{TIN98, HAR03}), albeit only discernible through a few faint knots. 

The {\it inner lobes} of Cen\,A are embedded in the thermal interstellar gas of NGC\,5128 (e.g. \citealp{FEI81}). They show up in radio and X-rays, with each lobe having a projected size of approximately 5\,kpc. The so-called {\it middle lobe}, regarded by \cite{MOR99} as an extension of the north-east inner lobe, has a size of $\sim30$\,kpc in projection and has no visible counterpart in the south. Based on their detection of extended thermal X-ray emission from this region, \cite{KRA09} interpret the northern middle lobe as an old structure that has recently become reconnected to the energy supply from the jet. Multiple age estimates for the individual lobes exist in the literature (\citealp{MOR99, SAX01, KRA03, CRO09, KRA09}), based on the dynamics. Each of the two {\it giant lobes} (Fig.\,\ref{fig:cena}) extends about 280\,kpc in projection and is positioned at a large angle to the inner lobes. \cite{HAR09} have derived spectral ages of the giant lobes of $\sim30$\,Myr.

X-rays produced by inverse-Compton scattering of cosmic microwave background photons are expected from the lobes of all FR\,I radio galaxies, though only a couple of examples (e.g. Centaurus\,B, {\it ASCA}, \citealp{TAS98}; NGC\,6251, {\it Suzaku}, \citealp{TAK12}) are known so far. The fields of view of {\it Chandra} and {\it XMM-Newton} are too small to map the giant lobes of Cen\,A and also the distribution of group gas surrounding them. {\it ASCA} did detect thermal emission from hot gas in a region associated with the northern giant lobe \citep{ISO01}, and most recently, \cite{STA13} have claimed thermal and non-thermal X-ray detection with {\it Suzaku} of parts of the southern giant lobe, though they do not detect inverse-Compton emission. {\it INTEGRAL} hard X-ray observations of Cen\,A's giant lobes by \cite{BEC11} are consistent with non-detection. The analysis of gamma-ray data by the \cite{FERMI10a} and \cite{YAN12} has shown that the gamma-ray radiation emanating from Cen\,A's giant lobes is of inverse-Compton origin, and that the lobes are particle dominated by a factor of a few. This is congruous with the results for the other two FR\,I galaxies resolved by {\it Fermi}-LAT, NGC\,6251 \citep{TAK12} and Centaurus\,B \citep{KAT13}.
 
In this paper we attempt to infer whether the properties of the giant lobes required for UHE particle acceleration are consistent with properties obtained from the constraints on energy input from the jet, on particle cargo and on dynamics. The paper is organised as follows. In Sect.\,\ref{sect:physicsjets}, we investigate physical conditions in Cen\,A's jets: we derive the jet power and place constraints on the energy density and entrainment rates. We focus on some of the fundamental physics -- the temperature of the thermal gas and the pressure and particle content -- of the giant lobes in Sect.\,\ref{sect:physicsGLs}, placing constraints on them using environmental information. We contrast the energy and particle supply through the jet with the pressure and energy content of the lobes and find that, to meet the pressure requirements, the protons must be unconventionally hot or relativistic. Finally, in Sect.\,\ref{sect:acceleration}, we explore the feasibility of stochastic UHECR acceleration models and the role of MHD turbulence and magnetic reconnection. Our main results and their implications are summarised, and prospects for current and future observations are drawn in Sect.\,\ref{sect:discussion}.

Spectral indices $\alpha$ are defined in the sense $S_{\!\nu}\propto\nu^{-\alpha}$, and particle indices $p$ as $n(E)\propto E^{-p}$.

\section{Physical conditions in the jets} \label{sect:physicsjets}

In this section we evaluate the degree of agreement between various jet power estimates, and investigate the energetics and mass loading of the jet. We compute the entrainment rates from hot gas and from stars within the jet and balance these values with requirements imposed by the pressure constraints on the lobes, which we will employ in Sect.\,\ref{sect:turbulence} in a consistency analysis when we consider stochastic particle acceleration in the large-scale lobes. 

\subsection{Jet power} \label{sect:jetpower}

\cite{CRO09} calculated a jet power for Cen\,A of $\sim1\times10^{43}$\,erg\,s$^{-1}$ based on the enthalpy of the southern inner lobe and its age estimated from the shock speed around the lobe of $\sim$2\,Myr, and an instantaneous jet power of $6.6\times10^{42}$\,erg\,s$^{-1}$ using the shock speed of 2600\,km\,s$^{-1}$. Applying the simple model of \cite{FAL99}, which relies on a relation between jet power and accretion disk luminosity, we infer from their Eq.\,20 a total jet power of $\sim1.6\times10^{43}$\,erg\,s$^{-1}$, adopting a core flux density of 3.9\,Jy at 8.4\,GHz \citep{MUL11}, 3.8\,Mpc for the distance \citep{HARR10}, a black hole mass of $5.5\times10^7\,M_{\odot}$ \citep{CAP09} and a jet viewing angle of 50$^\circ$ \citep{TIN98, HAR03}. Such a jet power is in reasonable agreement with the previous estimates, given the numerous assumptions involved. Note that increasing the viewing angle to 70$^\circ$ (e.g. \citealp{ JON96, TIN98}) in the Falcke \& Biermann model boosts the jet power to $\sim3.3\times10^{43}$\,erg\,s$^{-1}$. 

The \cite{FERMI10a} computed Cen\,A's kinetic jet power of $7.7\times10^{42}$\,erg\,s$^{-1}$ based on the synchrotron age of the giant lobes of 30\,Myr \citep{HAR09} and on an estimated total energy in both giant lobes of $\sim 1.5\times 10^{58}$\,erg. However, the imposition of the dynamical age of $\sim560$\,Myr (see Sect.\,\ref{sect:sizeandage}) would imply much lower average jet powers or a significantly higher total lobe energy. We will elaborate on this in Sect.\,\ref{sect:energetics}. 

The above methods, whose outcomes we express in terms of a single jet, all yield similar values for the jet power. Since the method based on the energy content of the giant lobes gives an estimate for the average past power of the jet, this means that if the jet power was higher in the past (as conjectured by, e.g., \citealp{SAX01}; \citealp{PRO10}), the jet activity in Cen\,A must have been intermittent.

\subsection{Constraints on energy density and magnetic fields} \label{sect:jetenergy}

Apart from the jet power needed to drive particular features of the source (as dealt with in Sect.\,\ref{sect:jetpower}), the jet power that {\it can} be carried by the known particle population can be estimated.

Considering the {\it inner jet}, that is the structure out to $\sim3$\,kpc of projected length, where we know the electron distribution reasonably well (see \citealp{HAR06}), we have a constraint on the bulk flow speed from the proper motion of the inner knots ($0.5c$, \citealp{HAR03}), and we are also confident that the magnetic field cannot be much lower than the equipartition value \citep{HAR11}. Do we require protons (thermal or relativistic) in order to transport
10$^{43}$\,erg\,s$^{-1}$, if we make the additional assumption that the jet is not
magnetically dominated?\footnote{It is unlikely that $U_{\rm B}>U_{\rm e}$ on these scales: models (e.g. \citealp{DRE02}) show that the conversion of Poynting to mechanically dominated jet occurs relatively close to the nucleus. Moreover, since the minimum energy condition is the equipartition condition, which we assume, it follows that a magnetically dominated jet must carry more energy, and thus if it is magnetically dominated the energy flux can be even higher without invoking protons, in general. So the values we obtain are lower limits if $U_{\rm B}>U_{\rm e}$, and the conclusions are unchanged.} Using the model of the jet from \cite{HAR11} with a single electron spectrum, the mean energy density of the jet (assuming equipartition between magnetic field and electrons only) is $U_{\rm j}=8.77\times10^{-11}$\,erg\,cm$^{-3}$. The total jet power follows from $P_{\rm j}= \Gamma^2\,4/3\,U_{\rm j}\,\pi r^2_{\rm j} \upsilon_{\rm j}$, where $\Gamma = (1-\beta^2)^{-1/2}$ is the bulk Lorentz factor and $r_{\rm j}$ is the cross-sectional radius of the jet. Assuming $r_{\rm j}$ = $18.8''$ (338\,pc) and $\upsilon_{\rm j}=0.5c$ (see above), 
\begin{figure}[!ht]
\begin{center}
\includegraphics[width=0.49\textwidth]{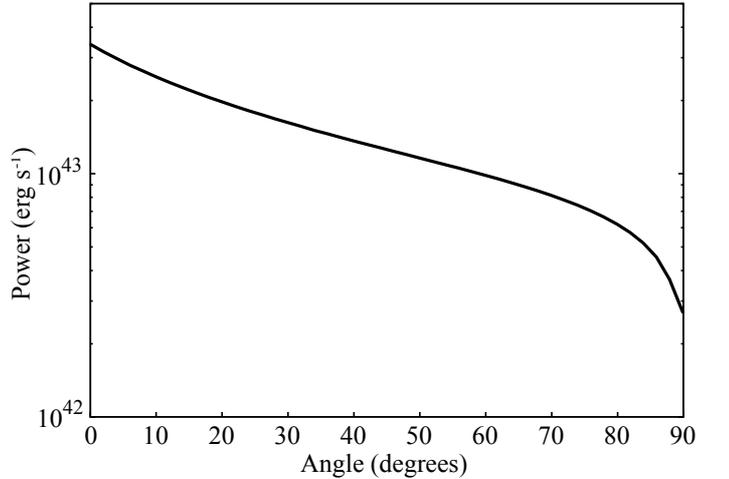}
\caption{Total jet power estimated with the formula in the text with the apparent jet speed fixed, as a function of the angle of the jet to the plane of the sky, using assumptions given by \cite{HAR11}.}\label{fig:jet-pow}
\end{center}
\end{figure}
for our preferred viewing angle of Cen\,A of $50^{\circ}$, the total jet power is $\sim1.3 \times 10^{43}$\,erg\,s$^{-1}$ (see Fig.\,\ref{fig:jet-pow}). This suggests that the energy transport can all be done by the known population of electrons with no requirement of energetically dominant protons.\footnote{A simple way of seeing that the effects of electron-positron annihilation are negligible in a purely leptonic jet on these scales is to consider the characteristic timescale for this process. Although the cross section for annihilation is energy-dependent (e.g. \citealp{MUR05}), the assumption that the cross section is the Thompson cross section and the particle speed the speed of light overestimates the cross section for all but the very lowest energies; thus, taking the timescale $\tau_{\rm ann} = n/({\rm d}n/{\rm d}t) = (n\sigma_{\rm T}c)^{-1}$ gives a conservative lower limit. If we take the rest-frame energy density for the jet given above, then the number density of relativistic electrons (with $n(\gamma) \propto \gamma^{-2}$, assuming $\gamma_{\rm min} = 1$ and $\gamma_{\rm max} = 10^8$) required is around $n_{\rm e,rel} = 3 \times 10^{-6}$\,cm$^{-3}$. Then $\tau_{\rm ann} \sim 530$\,Gyr, i.e. much larger than the Hubble time.} However, we cannot rule out the possibility that there is an energetically significant population of protons.

The magnetic field strength of the diffuse component of the jet is generally lower than that of the knots contained in the jet. \cite{GOO10} derived an equipartition value of the magnetic field of the inner knots in Cen\,A's jet (of which they associated the X-ray bright knots with shocks) in the range $\sim$\,220\,--760\,$\mu$G and a non-thermal knot pressure in the range $1.3\times10^{-9}-1.5\times10^{-8}$\,dyn\,cm$^{-2}$. It is probable that the knots are overpressured with respect to the diffuse component and also compared to the hot gas external to the jet, and we will return to pressure considerations in the next section.

\subsection{Matter densities, entrainment, temperature, and pressure balance} \label{sect:matter}

The idea that FR\,I jets start off purely leptonic (or at any rate very light) and become mass loaded through entrainment, thus allowing them to decelerate to sub-relativistic speeds, has received support from the work of \cite{LAI02a}. Mass loading may occur from hot and/or cold gas via the jet boundary ({\it external entrainment}), as a result of mass loss from stars within the jet volume ({\it internal entrainment}), or from the intermittency of the jets, i.e. when the jets restart ({\it prompt entrainment}). The prompt entrainment is expected to be insignificant compared to other contributions to the total entrainment, and we will disregard it in what follows. External entrainment from cold gas is difficult to quantify (e.g., it will not scale with distance in the same way, or with the external hot gas density as external entrainment from hot gas, see Sect.\,\ref{sect:external}). Various authors \citep{STI04, STR10, AUL12} have reported the existence of cold clouds along the jet axis but offer no direct evidence for jet-cloud interaction. We choose to also neglect the contribution from cold gas to the total entrainment.

There are no direct constraints on the thermal matter content of the Cen\,A jet. Observations of depolarization, or Faraday rotation deviating from a $\lambda^2$ law could, in principle, place constraints on the thermal electron number density of the knots \citep{BUR66}, when combined with known (\citealp{GOO10}, see Sect.\,\ref{sect:jetenergy}) constraints on the knot equipartition magnetic fields and knot sizes. \cite{CLA92} found no evidence for internal depolarisation, so that only upper limits on the thermal electron densities can be inferred. For the knots for which we have sufficient constraints, A1A, A1B and A1C, using equipartition fields of 693, 323 and 763\,$\mu$G \citep{GOO10}, this gives upper limits of $n_{\rm e,th}\sim3\times10^{-2}$, $\sim6\times10^{-3}$ and $\sim8\times10^{-2}$\,cm$^{-3}$, respectively. These are well above the densities that we estimate from entrainment.

\cite{DEY86}, \cite{BIC84, BIC94}, \cite{HEN87} and \cite{LAI02a} invoked external entrainment via a turbulent boundary layer, arising from Kelvin-Helmholtz (KH) instabilities at the jet boundary. Significant excursions from pressure equality between the jet and the surrounding ISM would disprove the existence of the KH instability along the jet boundary. The bright knots measured by Goodger et al. are very likely overpressured with respect to the bulk of the jet, and the minimum pressure values, $2.1\times10^{-11}-9.2\times10^{-11}$\,dyn\,cm$^{-2}$, determined by \cite{BUR83}, which are for larger areas and probably closer to being representative of a `mean jet' state, still suggest overpressure. However, the minimum jet pressure, adopting our energy density value from Sect.\,\ref{sect:jetenergy}, is $p_{\rm j}=U_{\rm j}/3\sim2.9\times10^{-11}$\,dyn\,cm$^{-2}$. The ISM pressure near the centre of the source is given in \cite{CRO09}: $\sim1.1\times10^{-11}$\,dyn\,cm$^{-2}$. Considering another location in the surrounding medium relatively close to the centre and using Kraft et al.'s (2003) values for density and temperature (see Sects.\,\ref{sect:external} and \ref{sect:sizeandage}), gives $\sim4.1\times10^{-11}$\,dyn\,cm$^{-2}$, which implies that the jet is not significantly overpressured with respect to the surrounding ISM if the jet pressure is close to its minimum value, and a KH instability along the jet is thus not ruled out. An actual proof of the existence of a turbulent jet boundary associated with KH instabilities would require detailed modelling including jet speeds and internal/external temperatures, and is beyond the scope of this paper, but we regard it as possible that thermal material could be entrained across a KH unstable jet boundary. However, there is no observational evidence for a KH instability associated with Cen\,A's jet.

\subsubsection{External entrainment from hot gas} \label{sect:external}

To estimate the level of external entrainment we will consider a jet that propagates in direct contact with the host galaxy's ISM. Our approach resorts to the use of simple scaling relations with position along the jet and we consider the results of \cite{LAI02a} for 3C\,31 to normalise the entrainment rate profile. We assume that the mass entrained per unit time for a section of jet of length $\Delta l$ scales according to the external gas density, the jet velocity, and the surface area of the jet segment, i.e.
\begin{equation}
\dot{M} = \Psi_0\, \rho_{\rm ext}(l)\, \upsilon_{\rm j}(l)\, r_{\rm j}(l)\, {\rm \Delta}l\,.
\end{equation}
Then the entrainment rate per unit length is
\begin{equation}
\Psi(l) = \Psi_0\, \rho_{\rm ext}(l)\, \upsilon_{\rm j}(l)\, r_{\rm j}(l)\,,
\end{equation}
where $\Psi_{0}$ is a normalisation factor, $l$ is distance along the jet,
$\rho_{\rm ext}(l)$ is the external mass density, $\upsilon_{\rm j}(l)$ the jet
velocity, and $r_{\rm j}(l)$ the jet radius at distance $l$.

We assume that the external thermal number density is described by a
beta model (e.g. \citealp{CAV76}):
\begin{equation}
n_{\rm th} = n_{\rm 0}\left[1 + (l / a)^2\right]^{-3\,\beta/2}\,,
\end{equation}
where $a$ denotes a scale radius, $\beta$ a slope parameter, and $n_{\rm 0}$ the external number density at $l=0$\footnote{A true pressure profile may deviate from the empirical isothermal model, but only markedly beyond large radii; this does not affect our calculations.}. For Cen\,A we adopt the beta model parameters from \cite{KRA03}: $a=0.5$\,kpc, $\beta=0.39$ and $n_{\rm p,0}=0.037$\,cm$^{-3}$.

We assume that the jet velocity is approximately constant over the
inner 2\,kpc of Cen\,A, and we assume that the jet radius $r_{\rm j}$ is
proportional to distance $l$, consistent with the fairly constant
opening angle observed over the region being considered. We can then
separate out the entrainment rate into normalisation and distance-dependent terms:
\begin{equation}
\Psi(l) = \Psi_{\rm norm}\, l \left[1 + (l/a)^{2}\right]^{-3\, \beta/2}\,.
\end{equation}

We determined the normalisation $\Psi_{\rm norm}$ by taking an estimate
of the entrainment rate in the middle of the flaring region of the
3C\,31 jet (based on the \citealp{LAI02a} model) -- choosing a
radius of 2\,kpc. Based on the jet geometry from high-resolution radio
data, and the extent of the X-ray jet in the two sources, we conclude
that the equivalent point within the flaring region of the Cen\,A jet
is at around 0.7\,kpc. We scale the entrainment rate at 2\,kpc for
3C\,31 ($\Psi = 4\times10^{22}$\,g\,s$^{-1}$\,kpc$^{-1}$) based on the
ratio of external density, jet velocity and jet radius at these
equivalent points as follows: $\rho_{\rm Cen A}/\rho_{\rm 3C\,31} = 0.33$,
$\upsilon_{\rm Cen\,A}/\upsilon_{\rm 3C\,31} \sim 1$, and $r_{\rm Cen\,A}/r_{\rm 3C\,31} = 0.08$. This leads to an estimated entrainment rate for Cen\,A at 0.7\,kpc of $\sim1.1\times10^{21}$\,g\,s$^{-1}$\,kpc$^{-1}$, and a normalisation for the entrainment rate profile of $\Psi_{\rm norm}\sim2.8\times10^{21}$\,g\,s$^{-1}$\,kpc$^{-2}$. Integrating the entrainment rate profile between
$l = 0$ and 3\,kpc implies a total entrainment rate of $\sim3.0\times10^{21}$\,g\,s$^{-1}$ ($\sim4.7\times10^{-5}$ M$_{\odot}$\,yr$^{-1}$). Assuming a mean particle mass of $0.6\,m_{\rm H}$, this is $\sim9.4\times10^{52}$\,particles\,yr$^{-1}$. Adopting a lifetime for the inner
lobes of 2\,Myr \citep{CRO09} gives a total mass injection of $1.9\times10^{35}$\,g ($\sim 94$\,M$_{\odot}$).

\subsubsection{Internal entrainment} \label{sect:internal}

We next calculate the entrainment from stars within the jet volume by estimating the total mass loss rate for the stellar population of Cen\,A, determining the fraction of this mass loss that occurs within the jet boundaries, and considering all this mass to be entrained. We assume that the contained stars are not affected by the jet plasma. We know (e.g. \citealp{KOM94, BOW96}) that mass loss in ellipticals is dominated by old-population stars, with $B$-band luminosity-to-mass-loss-rate ratio of $7.88\times10^{-12}$\,($L_{\rm B}$\,/\,$L_{{\rm B}{\odot}}$)\,$M_{\odot}$\,yr$^{-1}$ (\citealp{ATH02}, see also \citealp{SUL11b}). The apparent magnitude $m_{\rm B}$ of Cen\,A is 7.48 \citep{TUL88}, from which we derive the luminosity $L_{\rm B}\sim2.43\times10^{10}\,L_{{\rm B}{\odot}}$. Hence we obtain a total mass loss rate of $\sim0.19\,M_{\odot}$\,yr$^{-1}$. Adopting a spherically symmetric distribution, the fraction of the stars that lie within the jet is determined by the jet's solid angle: with Cen\,A's jet opening angle of 15$^{\circ}$ \citep{GOO10}, the solid angle is 0.054 steradians. This gives an entrainment rate from stars inside the jet of $\sim5.2\times10^{22}$\,g\,s$^{-1}$ ($8.2\times10^{-4} M_{\odot}$\,yr$^{-1}$) which converts to $1.6\times10^{54}$\,particles\,yr$^{-1}$. The internal entrainment is thus slightly larger than that from the hot ISM, which is not unexpected, as even for 3C\,31 where the galaxy environment has much less gas content, internal entrainment dominates the central parts. 

Reflecting the presence of a starburst around Cen\,A's nuclear region, we also consider the mass loss from O stars. We assume that the Cen\,A FIR luminosity of $3.4\times10^{43}$\,erg\,s$^{-1}$ ($\sim 8.9\times10^9$\,$L_{\odot}$) arises entirely from dust heated by type O stars, and that 50\% \citep{GIL07} of the output of these stars goes into heating the dust. Given a luminosity per star of $L = 3\times10^5\,L_{\odot}$ (for an O6.5 star, from \citealp{VAC96}), we have $5.9\times10^4$ such stars, each with a mass loss rate of $1\times10^{-6}\,M_{\odot}$\,yr$^{-1}$, for a total mass loss rate of $5.9\times10^{-2}\,M_{\odot}$\,yr$^{-1}\sim1.2\times10^{32}$\,g\,yr$^{-1}$. Adopting a spherically symmetric distribution and the solid angle as above, the entrainment rate from O stars inside the jet is $\sim1.6\times10^{22}$\,g\,s$^{-1}$ which converts to $5.0\times10^{53}$\,particles\,yr$^{-1}$. This is a factor of a few less than the rate invoking AGB stars, summing to an internal entrainment rate of $\sim6.8\times10^{22}$\,g\,s$^{-1}$. \cite{LAI02a} used a variant of those methods and derived for 3C\,31 an internal entrainment of $\sim4.9\times10^{23}$\,g\,s$^{-1}$.

\subsubsection{Pressure balance in the inner lobes}

We have good constraints on the internal pressure required in the
inner lobes of Cen\,A from the shock conditions of the south-west lobe
\citep{CRO09}. The total required internal pressure is
$p_{\rm SW-IL}=1.1\times10^{-10}$\,dyn\,cm$^{-2}$, where constraints from
the radio observations using the assumption of equipartition imply
that the relativistic pressure is $\sim 10\%$ of this, and hence $p_{\rm th}
\sim1\times10^{-10}$\,dyn\,cm$^{-2}$. The volume of the inner south-west lobe is $\sim 1.6\times10^{66}$\,cm$^{3}$ \citep{CRO09}, and so this implies a thermal particle number $N_{\rm th}\sim4.5\times10^{60}$ (assuming again a mean particle mass of $0.6\,m_{\rm H}$) and thus $n_{\rm th}\sim2.8\times10^{-6}$\,cm$^{-3}$. 

Given the total amount of entrained material, it would need to be heated to a temperature of $\sim2.6\times10^{11}$\,K to balance the pressure offset in the southern inner lobe\footnote{We are using `hot thermal plasma with a given $kT$' as a synonym for `plasma with mean particle kinetic energy of order $3kT/2$' and we are not particularly concerned with whether the particle energies have a Maxwell-Boltzmann distribution or not.}. Pressure support to Cen\,A's southern inner lobe supplied by either protons or lower-energy relativistic electrons has been proposed earlier by \cite{KRA03}; and in a more general case, \cite{BOH93}, \cite{CAR94}, \cite{HAR98a}, \cite{DUN04}, \cite{BIR08}, \cite{DIE08} and \cite{TAK12}, for example, argued for other particles in addition to relativistic electrons to match a missing pressure in other sources' lobes. We will handle pressure balance with regard to Cen\,A's giant lobes in Sect.\,\ref{sect:environmental}.

\section{Physical conditions in the giant lobes} \label{sect:physicsGLs}

Our objective is to assess the pressure and temperature of the thermal gas and particle content of the giant lobes, and to touch on energy distribution in the lobes. We begin by addressing several of the relevant morphological features of the lobes and by putting constraints on the lobe ages. Limits on lobe ages are required to evaluate the power of the jet that inflated the giant lobes (Sect.\,\ref{sect:energetics}) and to estimate the number of thermal protons entrained over the lifetime of the source (Sect.\,\ref{sect:abundances}).

\subsection{Size and age} \label{sect:sizeandage}

With the 1.4\,GHz images of \cite{FEA11} as a guide, and approximated as prolate spheroids with major and minor axes of $280\times170$\,kpc and $280\times190$\,kpc, the northern and southern giant lobes occupy volumes of $\sim1.2\times10^{71}$\,cm$^3$ and $\sim1.6\times10^{71}$\,cm$^3$ respectively\footnote{The sizes are approximate; the source lacks a clear boundary at 1.4\,GHz along most of its large-scale structure and there are uncertainties in the background/noise level. The southern lobe size includes the `gap' region.}. These sizes and volumes should be considered a lower limit, as the giant lobes may not lie perfectly in the plane of the sky. \cite{FEA11} have argued that the southern giant lobe is disconnected from the rest of the source and \cite{STE13} report a similar `gap' at 150\,MHz, yet we cannot exclude the possibility that the southern lobe is in the process of detachment, or is fully connected. If the lobe were connected, the fainter region could be explained because the synchrotron emission is a non-linear tracer of the underlying plasma ($j_{\rm sync}\propto U_{\rm e}^2\,B^2$), so that a small decrease in relativistic particles, magnetic field, or both can result in a large decrease in synchrotron brightness. 

\cite{HAR09} derived synchrotron ages of $t_{\rm sync}\sim25$\,Myr and $\sim27$\,Myr for, respectively, Cen\,A's northern and southern giant lobes. If we proceed with calculating the sound-crossing timescale 
\begin{equation}
t_{\rm cs}=R/c_{\rm s}
\end{equation}
(see, e.g., \citealp{BIR04, DUN04, DUN05, MCN07}), where $R$ is the distance to the lobe edge and $c_{\rm s}=\sqrt{\,\gamma k T/\mu m_{\rm H}}$ the local sound speed, by using the well-determined \citep{KRA03} temperature of $kT\sim0.35$\,keV (i.e. $\sim4.1\times10^6$\,K) and $R=280$\,kpc, the ratio of specific heats $\gamma=5/3$ and the mean particle mass $\mu=0.62$, we get $t_{\rm cs}\sim920$\,Myr. However, the value of 0.35\,keV applies to the external medium on the scales of the inner lobes. The temperature of the plasma into which the giant lobes are expanding over most of their lifetime is currently inaccessible, but could easily be higher, and we estimate it in Sect.\,\ref{sect:environmental} to be in the range $kT = 0.7-1.5$\,keV (i.e. $\sim8.1\times10^6-1.7\times10^7$\,K), resulting in a more likely sound-crossing timescale in the range $\sim440-645$\,Myr. The uncertainty in the projection angle of the giant lobes makes those values a lower limit.

The discrepancy between the figure for the synchrotron age and the sound-crossing timescale is not surprising: using spectral breaks (and inferred synchrotron age) to estimate the physical age of a source can be misleading, especially if {\it in situ} re-acceleration is taking place (e.g. \citealp{PAC76, ALE87, EIL94, EIL97, BIC95, FER98, KAT13}), as envisaged for the giant lobes of Cen\,A (\citealp{FERMI10a, FEA11, STA13, STE13}, and this paper). 

In addition, given that one of the giant lobes is considered to be disconnecting/disconnected -- although this is not a strict requirement; the buoyant force on the outer ends of the giant lobes just needs to be larger than any other force -- we compute the buoyancy age for the lobes. The buoyancy age is given as 
\begin{equation}
t_{\rm \:\!buoy}=R_{\rm dist}/\upsilon_{\rm buoy}\,, \label{eq:buoy}
\end{equation}
where we take $R_{\rm dist}$ to be the distance from the AGN core to the edge of the giant lobe\footnote{Simple treatments of rising bubbles track the position of the lobe centre. However, those treatments consider lobes which are much smaller than their height (or distance from the mass centre), so the bubbles are essentially test particles.} and the buoyancy velocity $\upsilon_{\rm buoy}=\sqrt{\,2gV/S C_{\rm D}}$, in which $V$ stands for the lobe volume and $S$ for its cross-sectional area (e.g. \citealp{FAB95, CHU01, BIR04, DUN04}). We assume a drag coefficient $C_{\rm D} =0.75$ as in \cite{CHU01}. Since the buoyant blob has to be small compared to the scale over which gravity varies (e.g. \citealp{FAB95}), Eq.\,\ref{eq:buoy} is not appropriate for the entire vertical, long giant lobe. Adopting a spherical bubble with radius of 80\,kpc (i.e., an approximate radius of the giant lobes towards their outer edges) which gives us a volume of $\sim6.3\times10^{70}$\,cm$^3$, and taking gravitational acceleration\footnote{We assume that $g$ is not too variable over the region.} $g$ resulting from an enclosed gravitating mass of $M_{\rm grav}(<R_{\rm dist})\sim1.4\times10^{46}$\,g (based on the mean of the orbital and virial masses from \citealp{KAR07} determined using the harmonic radius of 192\,kpc), results in $\upsilon_{\rm buoy}\sim4.9\times10^7$\,cm\,s$^{-1}$ and hence, appealing to the adopted lobe edge $R=280$\,kpc, in $t_{\rm \:\!buoy}\sim560$\,Myr. This puts the buoyancy age and sound-crossing times close, as expected from basic physics (virial theorem). However, the above derived buoyancy age is probably a conservative estimate, reflecting the dependence on other physics (ram pressure or high internal pressure exceeding the buoyancy force during some period of the lobe growth will make the lobes younger), on the projection (an inclination angle would make the lobes older), and, to a lesser extent, of the drag coefficient on the Reynolds number (for the latter dependence see \citealp{FAB95}). No independent age estimates are available from, e.g., proper motion from the edge-like features (the {\it wisps}, see \citealp{FEA11}). 

Dynamical age estimates for FR\,I sources are scarce, and we can only compare with 3C\,31 for which \cite{PER07} derived $>\!100$\,Myr, and with J1453+3308 (a FR\,I/FR\,II source) for which two estimates exist in the literature: 215\,Myr \citep{KAI00a} and $\leq134$\,Myr \citep{KON06}.

\subsection{Matter densities, pressure and temperature} \label{sect:environmental}

Based on Faraday RMs and linearly polarised intensities of background sources, and adopting the path length through the lobes as 200\,kpc, a magnetic field strength in the lobes of 1.3\,$\mu$G and assuming no field reversals along the line of sight throughout the lobes, \cite{FEA09} placed a limit on the volume-averaged thermal electron number density of Cen\,A's giant lobes of $n_{\rm e,th}\lesssim5\times10^{-5}$\,cm$^{-3}$. Using $B=0.9$\,$\mu$G (from the \citealp{FERMI10a}, see Sect.\,\ref{sect:energetics}) this limit becomes $n_{\rm e,th}\lesssim7\times10^{-5}$\,cm$^{-3}$. There are claims (\citealp{STA13, SUL13}) that the thermal particle number density of Cen\,A's giant lobes may be as high as $\sim1\times10^{-4}$\,cm$^{-3}$; however, this is inconsistent with observations of X-ray surface brightness decreases (cavities) in other radio galaxies. Moreover, Stawarz et al.'s thermal X-ray detection may be interpreted in terms of Galactic foreground emission (in their Fig.\,7, an extended region of X-ray emission contiguous with Galactic emission appears to lie on top of the regions they use).

Let us now derive constraints on the internal pressure, temperature and particle content of the giant lobes, considering environmental constraints. \cite{KAR07} give a total mass for the Centaurus group of $(7\,-\,9)\times10^{12}\,M_{\odot}$. \cite{SAN03} show well-constrained $M_{\rm tot}$ -- $T_{\rm X}$ and $L_{\rm X}$ -- $T_{\rm X}$ relations for a sample of (relaxed) galaxy groups and clusters (with $\sim12$ sample members having temperatures below 2\,keV, so in the group regime). Using their $M_{\rm tot}$ -- $T_{\rm X}$ relation, the above mass for the Centaurus group implies an X-ray gas temperature between 0.7 and 1.5\,keV, which in turn implies an X-ray luminosity between 10$^{42}$ and 10$^{43}$\,erg\,s$^{-1}$.

These X-ray luminosities are very typical of radio-galaxy group-scale environments (e.g. \citealp{CRO08}). If we consider systems of similar luminosity for which the external pressure profiles have been mapped, there are $\sim5$ systems in \cite{CRO08}: NGC\,6251, NGC\,1044, 3C\,66B, NGC\,315 and NGC\,4261. Their thermal pressures at $\sim100$\,kpc range from $(1-5)\times10^{-12}$\,dyn\,cm$^{-2}$, and at 300\,kpc (i.e. comparable to the distance to the outer edge of the Cen\,A giant lobes) from $(1-10)\times10^{-13}$\,dyn\,cm$^{-2}$.

If we take the median values of the external pressure at these radii, we have some plausible (if not well-constrained) estimates of the required internal pressure, assuming that the lobes are within a small factor of pressure balance.

The {\it Fermi}-LAT inverse-Compton analysis \citep{FERMI10a} gives a relativistic pressure equal to $p_{\rm rel} = 5.6\times10^{-14}$\,dyn\,cm$^{-2}$ (northern giant lobe) and $2.7\times10^{-14}$\,dyn\,cm$^{-2}$  (southern giant lobe). The {\it Fermi}-LAT results therefore imply ratios at 100\,kpc ($\sim$giant lobe midpoints) of $p_{\rm tot}/p_{\rm rel}$ = 18\,-\,90 (northern giant lobe) and 37 - 185 (southern giant lobe), which means that the pressure would have to be dominated by non-radiating particles -- thermal or not. 

For these ratios, if we assume the pressure is dominated by thermal particles, we can infer the density of thermal particles in the
giant lobes for various electron temperature assumptions. If we assume an internal temperature $T\sim10^7$\,K, as claimed by \cite{ISO01} on the basis of detected diffuse X-ray emission, then $n_{\rm e,th}\sim1\times10^{-3}$\,cm$^{-3}$, which is inconsistent with the Faraday rotation limits of \cite{FEA09} and also our revised value. Adopting instead the limit of $n_{\rm e,th}\lesssim7\times10^{-5}$\,cm$^{-3}$, and taking a mean thermal pressure of $1.5\times10^{-12}$\,dyn\,cm$^{-2}$ following our modelling above, then we find a lower limit to the temperature of $\sim1.6\times10^8$\,K.

There is reason to favour such a high temperature for the thermal gas, based on observations of cavities associated with radio lobes in other galaxies (e.g. \citealp{BIR04}), which imply that the temperature of thermal material contained within lobes must be sufficiently high to provide the required pressure with comparatively low density gas so as to minimize thermal bremsstrahlung. Limits on the temperature of this gas have been obtained by several authors: e.g. $>\!\!15$\,keV (Hydra\,A, \citealp{NUL02}), $>\!\!20$\,keV (Abell\,2052, \citealp{BLA03}), $>\!50$\,keV (Perseus\,A, \citealp{SAN07}).

If our density estimates from entrainment (Sect.\,\ref{sect:abundances}) are correct then the
thermal material must be very hot indeed. Our constraint on $n_{\rm p,th}$ (Sect.\,\ref{sect:abundances}) combined with our thermal lobe pressure of $\sim1.5\times10^{-12}$\,dyn\,cm$^{-2}$, gives a temperature estimate of $\sim2.0\times10^{12}$\,K. This means that if the protons provide the missing pressure that we claim in Sect.\,\ref{sect:environmental} (but note that there is no direct observational evidence for the statement that this is the amount of pressure we need) then they must be very hot. However, to put this in context, $k\times10^{12}$\,K is only $\sim10\%$ of the proton rest mass, so the protons have the same energy as $\gamma\sim250$ electrons, which are abundant in the giant lobes. Hence, there is a self-consistent model in which the entrained protons from Sects.\,\ref{sect:external} and \ref{sect:internal} are heated to $\sim10^{12}$\,K (i.e. mildly relativistic) and then provide the missing pressure while still allowing UHECR acceleration; and that, even ignoring the UHECR models, this is the {\it only} model that explains what would otherwise be a discrepancy between the number of protons we expect to be entrained and the number we need in the inner/giant lobes (Sects.\,\ref{sect:external} and \ref{sect:internal}). Note that in case of jet intermittency over the lifetime of the AGN, even less material may be available to match the pressure. 

In summary, there is a self-consistent model where the external pressures are comparable to well-studied FR\,I radio galaxy environments, the thermal particle density is well below the limits from radio polarisation, and the thermal material in the lobes is hot.

\subsection{Magnetic field estimates and global energetics} \label{sect:energetics}

The use of radio observations in conjunction with X-ray or gamma-ray data permits a direct appraisal of the magnetic field strength. Observations at gamma-ray frequencies are preferable as these are not hampered by thermal bremsstrahlung as {\it in casu} X-ray bands. Based on radio data presented by \cite{HAR09} and the detection of soft gamma rays from the giant lobes of Cen\,A \citep{FERMI10a}, the latter authors obtained a field strength of $B$ = 0.89\,$\mu$G and 0.85\,$\mu$G for the entirety of respectively the northern and the southern giant lobes. Their computed energy density ratio $U_{\rm e}/U_{\rm B} = 4.3$ for the northern and 1.8 for the southern giant lobe indicates modest electron pressure dominance, which is analoguous to the values for the lobes of, e.g., Centaurus\,B ($U_{\rm e}/U_{\rm B}\sim4.0$; \citealp{KAT13}). As a cautionary note, since the particle index $p_1$ used by the \cite{FERMI10a} for the southern giant lobe is rather low and gives rise to a significantly lower value of $U_{\rm e}/U_{\rm B}$ than found elsewhere in the source, we use the $U_{\rm e}/U_{\rm B}$ value for the northern lobe in the Alfv\'en speed derivation in Sect.\,\ref{sect:consequences}.

Using the {\it Fermi}-LAT Table S1 entries, we calculate the relativistic electron number densities for the four giant lobe sectors (defined in \citealp{HAR09}): sector 1, $n_{\rm e,rel}\sim7.9\times10^{-9}$ cm$^{-3}$; sector 2, $n_{\rm e,rel}\sim1.5\times10^{-8}$ cm$^{-3}$; sector 4, $n_{\rm e,rel}\sim1.0\times10^{-11}$ cm$^{-3}$ and sector 5, $n_{\rm e,rel}\sim2.9\times10^{-10}$ cm$^{-3}$; these are notably different from one another. The reason for the widely varying electron number densities are presumably the substantial variations in the synchrotron surface brightness across these regions. \cite{YAN12} updated the inverse-Compton analysis based on a {\it Fermi}-LAT data set of three times the size of the \cite{FERMI10a} analysis, in which they confirm the earlier results for $U_{\rm e}/U_{\rm B}$.

The {\it Fermi}-LAT analysis and their resulting total energy $E_{\rm tot}\sim7.3\times10^{57}$\,erg per giant lobe only considers electron-positron plasma. In our picture the total energy supply must have been provided by the jet; significant contribution of non-radiating particles to the giant lobe pressure (as we envisage in the preceding section) increases their inferred value for $E_{\rm tot}$. In Sect.\,\ref{sect:consequences} we derive a proton pressure of $p_{\rm p}\sim1.5\times10^{-12}$\,dyn\,cm$^{-2}$. The total energy of an individual giant lobe, taking the mean lobe volume of $1.4\times10^{71}$\,cm$^3$, is then $E_{\rm tot}=4 p_{\rm e+p+B}\,V_{\rm l}\sim8.8\times10^{59}$\,erg; as a corollary, the power estimate of the jet that inflated the giant lobes increases (adopting our buoyancy age\footnote{The sound-crossing timescale is not as useful an age estimate since it is hard to justify physically for the large-scale lobes.} of 560\,Myr) to $\sim5.0\times10^{43}$\,erg\,s$^{-1}$. In this picture, the agreement between the jet power derived for the giant lobes by the \cite{FERMI10a} and the estimate made in the inner jets is coincidental, since we would argue that both the energy content and the lobe age that they estimate are too low.

The minimum pressure analyses in the literature make the conservative assumptions that there are no relativistic baryons, and that the emitting volume is uniformly filled. For the giant lobes, we are not using an equipartition field but one determined by the \cite{FERMI10a} using inverse-Compton, and the traditional parameter $k$, i.e. the ratio of the total particle energy density to that in relativistic electrons, does not affect such measurements. The plasma filling factor $f$ does, but in a slightly more complicated way (see discussion in \citealp{HAR00}): only if the electrons had a very low $f$ could such a model give much larger pressures than the inverse-Compton value, and such a model also requires a high-pressure non-radiating fluid.

\subsubsection{Electron-positron and proton content} \label{sect:abundances}

So far, we have made no assumption about whether the relativistic material is electron-positron or electron-ion plasma, excepting the jet on the smallest scales (Sect.\,\ref{sect:matter}) for which we assume a (nearly) pure electron-positron plasma. All of our estimates of energy densities ($U_{\rm e,rel}$) and pressures ($p_{\rm e,rel}$) in the radiating particles account for positrons. 

Since cooling is energy-dependent, the relativistic electron-positron population on the smallest scale is expected to remain non-thermal while propagating along the jet. The electrons do not approach sub-relativistic energies for realistic lifetimes.

We associate the thermal component of the jet/lobes with thermal electrons and positrons from external and internal entrainment. Note that by `thermal protons' we mean `thermal protons and electrons' just as with `relativistic electrons' we mean `relativistic electrons and positrons'. It is reasonable to expect that the entrained electrons and positrons will behave in the same way as the thermal protons, but we do not know whether either species is heated or turned into a non-thermal distribution via particle acceleration processes; no statement is possible on the fraction of the entrained electrons and positrons which becomes non-thermal while being transported along the jet.

Pure thermal electron-positron plasma would be excluded by a Faraday rotation detection, yet we have no firm constraints on this in either the jet or the giant lobes. The thermal electrons and positrons are not required for energy transport in the jet (see Sect.\,\ref{sect:jetenergy}), and their temperature in the jet is not constrained by any observation we can make.

Considering the total entrainment rate of $7.1\times10^{22}$\,g\,s$^{-1}$, the buoyancy age of 560\,Myr, and the average volume of the giant lobes, $1.4\times10^{71}$\,cm$^{3}$, we obtain a ballpark figure for the thermal proton number density in the giant lobes of $n_{\rm p,th}\sim5.4\times10^{-9}$\,cm$^{-3}$. We expect on charge balance grounds that $n_{\rm e,th}\simeq n_{\rm p,th}$. The consistency with the upper limit on $n_{\rm e,th}$ (Sect.\,\ref{sect:environmental}) is trivially true, and we know by the lack of low-frequency Faraday depolarisation (e.g. \citealp{WIL78b, JAG87}) and by direct observation of cavities of other radio galaxies (e.g. \citealp{BIR04}) that there is at least several orders of magnitude difference between the internal and external densities. For the Centaurus intragroup medium, \cite{SUL13} have suggested $n_{\rm th}\sim1\times10^{-4}$\,cm$^{-3}$.

Therefore, we will calculate the Alfv\'en speed (Sect.\,\ref{sect:consequences}) on the assumption of only two components of the (non-magnetic) energy density and pressure: a relativistic component whose energy density is constrained by the synchrotron and inverse-Compton observations, and a thermal electron/proton plasma whose density and temperature are constrained by the external pressure and entrainment arguments given above and in Sect.\,\ref{sect:environmental}.

\section{(UHE)CR acceleration} \label{sect:acceleration}

In this section we investigate whether the models we have presented for the particle content and energetics of the giant lobes of Cen\,A are consistent with a scenario in which UHECR are accelerated in the giant lobes. Specifically, in Sect.\,\ref{sect:UHECRs}, we derive UHECR power, in Sect.\,\ref{sect:enrichment}, we assess abundances in the giant lobes and the prevalent particle species available for acceleration. In Sect.\,\ref{sect:turbulence}, we discuss magnetic field fluctuations in the plasma of the giant lobes and their role in lepton and hadron acceleration along with possibilities for turbulent input, maintainance, heating and dissipation, and we discuss scenarios in which particles are subject to a hybrid mechanism invoking magnetic reconnection that could provide seeds for further energisation in the lobes.

\subsection{UHECR luminosity and power} \label{sect:UHECRs}

The \cite{ABR10b} measured 13 events $\ge\!55$\,EeV within a radius of $18^{\circ}$ from Cen\,A, with associated energies totalling 888\,EeV. With the distance to Cen\,A of 3.8\,Mpc, and correcting for PAO's directional exposure (isotropic exposure of 20\,370\,km$^2$\,yr\,sr, exposure to Cen\,A of 3095\,km$^2$\,yr), the UHECR luminosity is $\sim2.5\times10^{39}$\,erg\,s$^{-1}$. Assuming a particle number flux spectrum as $E^{-2}$, this gives a luminosity of ${\rm d}L/{\rm d}E=k E^{-1}$, where $k$ is a normalisation constant: $k\!=\!L_{\rm UHECR}/(1/E_{\rm min,norm}\!-\!1/E_{\rm max,norm})\!\!=\!\!2.5\times10^{39}$\,erg\,s$^{-1}$/(1/55\,EeV$-$1/84\,EeV) $\sim 1.33\times10^{47}$\,erg$^2$\,s$^{-1}$. For the high-energy cutoff we adopt the highest energy of the events reported in the \cite{ABR10b}, $E_{\rm max}=142$\,EeV. The low-energy cutoff is uncertain, but the total energy depends only weakly on this value: we use $E_{\rm min}=m_{\rm p}c^2\sim938$\,MeV. We find that the ratio between $L_{\rm UHECR}$ and the power put into cosmic rays in total is $\sim25$ (i.e., protons $\sim27.1$, $\chem \iso{7}{}Li\sim25.1$, $\chem \iso{16}{}O\sim24.2$, $\chem \iso{56}{}Fe\sim22.9$). This means about $6\times10^{40}$\,erg\,s$^{-1}$ goes into CRs of all types, which is reasonable given the jet power, and it leaves room for energy to go into other particle populations.

\subsection{Enrichment} \label{sect:enrichment}

Hardcastle's (2010) estimate of $\sim$1 iron nucleus per $10^5$ protons in the giant lobes refers to the hot ISM and so only considers external entrainment. AGB stars and O stars produce intermediate-mass nuclei (see in this context e.g. \citealp{KAR10a}) which are incorporated through internal entrainment, however they do not produce the isotope $\chem \iso{56}{}Fe$, and solely inject $\chem \iso{56}{}Fe$ at the initial abundances, which are insignificant. Therefore, effectively, internal entrainment will entrain material which will be enriched with the lighter intermediate elements  (mainly the CNO nuclei) with respect to the externally entrained material. The acceleration of these nuclei in the giant lobes could increase the UHECR flux. Hence, the thermal material in the giant lobes may well be enriched in light elements from stellar winds from stars within the jet; this alleviates the objection of \cite{LIU12} to the giant lobes as a source of UHECRs and may help to explain the PAO composition results (as per, e.g., the \citealp{PAO11a, PAO13}). A mounting body of reasoning (e.g. \citealp{FAR08, BIE12}), based on disparate analyses, is beginning to support the dominance of light to intermediate nuclei.

\subsection{Turbulence and Alfv\'enic acceleration} \label{sect:turbulence}

Giant lobes of FR\,I sources are expected to be turbulent to some degree (e.g. \citealp{CAV78}). \cite{JUN93} alluded to the possibility that the chaotic behaviour of polarisation associated with the southern giant lobe of Cen\,A may well be due to internal turbulence, and \cite{FEA09} found a depolarised signal and RM fluctuations intrinsic to the southern giant lobe of Cen\,A that they ascribe to turbulent magnetised plasma inside the lobe. 

\cite{EIL89a} argued that MHD turbulence will engender fluctuations in the total radio intensity. Standard fluid turbulence studies find an outer scale (i.e. driving scale) of the turbulence which is some fraction of the width of the flow; 1/7 `classically', and no larger than some large fraction ($\sim1/3$) of the size of the system. Observations of synchrotron structure at 1.4\,GHz on nearly the largest lobe scale \citep{FEA11} require that there must be magnetic field structure on these scales, and so are consistent with the idea that there is turbulence on these scales too. The Cen\,A literature shows a spread of driving scales, or maximum eddy sizes (assuming forward cascade), in these lobes, $\lambda_{\rm max}=10$\,kpc (\citealp{SUL09, SUL11a}) up to $\lambda_{\rm max}=100$\,kpc \citep{HAR09}. The smallest driving scale, 10\,kpc, seems invalidated by the 1.4\,GHz observations \citep{FEA11} at $\sim50''$ angular resolution -- among other the filament sizes. In our model for particle acceleration, power in turbulence on scales $\ga 100$\,kpc would have little or no effect on particle acceleration even if it were present, both because the scattering on Alfv\'en waves would be inefficient\footnote{This model (see Sect.\,\ref{sect:consequences} for a detailed discussion) requires a field of quasi-isotropic Alfv\'en waves at the resonant scale, not merely one or two eddies just fitting into the lobe.} and because the gyroradius of such particles would approach the size of the lobes, and so it is reasonable to adopt a scale of 100\,kpc as a hard upper limit. Since the largest scale of coherent filamentary structures in the lobes is around 30\,kpc, we adopt this as the minimum possible driving scale for turbulence. As the true driving scale for turbulence must lie within this range (though consideration of the hydrodynamics might favour lower values), we discuss all parameters that depend on this scale using the two extreme values of $\lambda_{\rm max}$.

\subsubsection{Turbulent input, maintenance and dissipation} \label{sect:input}

On energetic grounds, the turbulent input most likely originates from the jet. The question is whether either or both giant lobes are still connected to the energy supply, and, obviously, if we require UHECR acceleration by this method, what the timescale for decay of turbulence is. We deal with the latter by writing $\tau_{\rm t}=\zeta \lambda_{\rm max}/\upsilon_{\rm t}$, for which we adopt a numerical factor $\zeta=2$ (e.g. \citealp{MAC98} and references therein), the driving scale $\lambda_{\rm max}=30-100$\,kpc, and the turbulent speed $\upsilon_{\rm t}=0.063c$ (see below). This gives us $\sim3-10$\,Myr, which means that, after the energy supply by the jet to the giant lobes has ceased, in the aftermath of the old jet activity another few Myr are available for turbulent acceleration. If the current jet is the one powering the inner lobes, then it cannot be older than the lobes it has formed, and we have a robust constraint on the age of those (2\,Myr, \citealp{CRO09}).

An attractive alternative might be turbulent input from explosions of massive stars in the starburst at Cen\,A's core. Taken at face value, the starburst may be both too far from the lobe and energetically inadequate -- multiplying the supernova rate in Cen\,A by the available mechanical energy in supernovae ($1\times10^{51}$\,erg) suggests that one supernova would be needed every $\sim3$ years to compare to the jet. This does not tally with observations given that the last supernova seen in Cen\,A was SN\,1986\,G \citep{EVA86, CRI92}, not to mention the difficulty of getting that energy to impact exclusively on the lobe. Admittedly, it is somewhat unclear whether SN\,1986\,G is a firm benchmark as one might have missed supernovae going off in the dusty regions of the galaxy. Even so, the rate would have to be several orders of magnitude higher than observed to deal with the coupling to the giant lobe scale. We therefore view the starburst activity as a very small, if not non-existent, additional contribution to the turbulent input to the lobe.

We also disregard turbulent input from galaxy mergers: the physical picture is more relevant for, e.g., intracluster media (see \citealp{PLA12}), let alone the timescale of the last merger associated with NGC\,5128 which is estimated to be significantly larger than a few Myr (e.g. \citealp{REJ11}).

If we suppose that the magnetic field is maintained by a turbulent MHD dynamo, then the proposition also implies $\rho\upsilon_{\rm t}^2\simeq B^2/8\pi$. Using $n_{\rm p,th} = n_{\rm e,th}=5.4\times10^{-9}$\,cm$^{-3}$ (see Sect.\,\ref{sect:abundances}) and $B=0.9$\,$\mu$G (from \citealp{FERMI10a}) puts a constraint on the turbulent speed of $\sim1.9\times10^9$\,cm\,s$^{-1}$ ($\sim0.063c$). Reassuringly, this is close to the Alfv\'en speed in the lobes as we derive in the subsequent section, therefore the requirement $\upsilon_{\rm t}\sim\upsilon_{\rm A}$, as applies in the turbulent MHD dynamo, is satisfied.

In a driven lobe, the turbulent energy is dissipated at a rate $\dot{\varepsilon}=U_{\rm B}\,\upsilon_{\rm t}/\,\zeta \lambda_{\rm max}$. Using the above values $\upsilon_{\rm t}\sim0.063c$, $\zeta=2$ and the driving scale $\lambda_{\rm max}=30-100$\,kpc gives us a turbulent dissipation of $\sim9.9\times10^{-29}-3.3\times10^{-28}$\,erg\,cm$^{-3}$\,s$^{-1}$. Multiplying this by the mean giant lobe volume ($1.4\times10^{71}$\,cm$^3$), we obtain a total power dissipation of $\sim1.4\times10^{43}-4.6\times10^{43}$\,erg\,s$^{-1}$, which is close to the power of the former jet that we have estimated in Sect.\,\ref{sect:energetics} based on pressure arguments. Thus this level of turbulence could plausibly have been driven by the pre-existing jet.

\subsubsection{Turbulent heating} \label{sec:heating}

We assume that the predominant fraction of the energy of the turbulence goes into heating the thermal plasma and a lesser portion into particle (re)acceleration, which is our best guess from our understanding of the physics. For $p_{\rm th} = 1.5\times10^{-12}$\,dyn\,cm$^{-2}$ and $B=0.9$\,$\mu$G (see Sects.\,\ref{sect:environmental} and \ref{sect:energetics}), the {\it plasma} $\beta = p/(B^2/8\pi)\sim47$. This may not have a major effect on the properties of the Alfv\'en waves (see \citealp{FOO79}) but as \cite{HOW10} shows it might affect the efficacy of particle heating through interactions with the MHD turbulence, with the merit that for high {\it plasma} $\beta$ media it operates more efficiently for hadrons than for leptons.

\subsubsection{Consequences for UHECR production in the lobes} \label{sect:consequences}

Following earlier works (e.g. \citealp{LAC77, EIL79, HEN82, SUL09}), we consider a model in which particle acceleration in the giant lobes is provided by a turbulent field of resonant Alfv\'en waves. In this process, a particle interacts via the cyclotron resonance with waves of wavelength comparable to the particle's gyroradius: $\lambda_{\rm res}\sim r_{\rm g}$, with $r_{\rm g} = \gamma m c^2 / Z e B$.\footnote{Particle acceleration by long-wavelength ($\lambda \gg r_{\rm g}$) magnetosonic turbulence has also been suggested by several authors (e.g. \citealp{KUL71, ACH81}). However, the stronger damping likely for magnetosonic waves (e.g. \citealp{EIL79, BIC82}) makes this process less credible to explain UHECR acceleration in Cen\,A. We note that the term `second order Fermi acceleration' has been used inconsistently in the literature, sometimes referring to all stochastic MHD processes, other times referring specifically to acceleration by magnetosonic turbulence. We therefore avoid using the term, and just refer to stochastic Alfv\'enic acceleration.} Because the Alfv\'en waves exist only with wavelengths up to the maximum turbulent scale in the source, there is a maximum particle energy which can resonate with the Alfv\'enic turbulence. Higher-energy particles, which cannot resonate with any Alfv\'en wave, are energised much less efficiently, and thus are probably not relevant for UHECR acceleration in the system. Taking $\lambda_{\rm max}=r_{\rm g}=30-100$\,kpc and $B=0.9\mu$G, and considering protons, we get $\gamma\sim2.7\times10^{10}-8.9\times10^{10}$, which translates to a proton energy of $\sim25-83$\,EeV. By way of comparison, for $\chem \iso{12}{}C$ this translates to $\gamma\sim1.3\times10^{10}-4.5\times10^{10}$ and thus an energy of $\sim150-500$\,EeV, for $\chem \iso{16}{}O$ to $\gamma\sim1.3\times10^{10}-4.5\times10^{10}$ and an energy of $\sim200-666$\,EeV and for $\chem \iso{56}{}Fe$ to $\gamma\sim1.2\times10^{10}-4.1\times10^{10}$ and an energy of $\sim650-2165$\,EeV. Requiring a lower limit of 55\,EeV (i.e. the low-energy threshold used for UHECRs by the \citealp{ABR10b}) allows lithium and heavier nuclei to `fit' (i.e. resonate with) the turbulent spectrum in the lobes if the driving scale is 30\,kpc. A 100\,kpc driving scale would also allow UHE protons to `fit'. 

In the model we are discussing, the important components of the energy density and pressure in the giant lobes are the thermal protons, the relativistic electrons, and the magnetic field, where those terms have the meanings defined in Sect.\,\ref{sect:abundances}, and so the Alfv\'en speed can be calculated as follows:
\begin{equation}
\upsilon_{\rm A} = \frac{c}{(1 + (U_{\rm p,th} + p_{\rm p,th} + U_{\rm e,rel} + p_{\rm e,rel})/(2 U_{\rm B}))^{1/2}}\,, \label{eq:alfven}
\end{equation}
with $U_{\rm p,th}=n_{\rm p}\,m_{\rm p}\,c^2\sim8.1\times10^{-12}$\,dyn\,cm$^{-2}$, and a pressure stored in thermal protons of $p_{\rm p,th}\sim1.5\times10^{-12}$\,dyn\,cm$^{-2}$. Supplemented by $U_{\rm e,rel}=4.3U_{\rm B}\sim1.4\times10^{-13}$\,dyn\,cm$^{-2}$ and $p_{\rm e,rel}\sim4.6\times10^{-14}$\,dyn\,cm$^{-2}$ (based on the \citealp{FERMI10a}), this results in an Alfv\'en speed of $\sim2.4\times10^9$\,cm\,s$^{-1}$ ($\sim0.081c$).

The resonant acceleration time for a particle of energy $\gamma$ can be approximated, to within a factor of order unity, as 
\begin{equation}
\tau_{\rm res}\simeq\frac{\gamma m c}{Z e B}\frac{c^2}{\upsilon_{\rm A}^2}\frac{U_{\rm B}}{U_{\rm res}}\,, \label{eq:resonant}
\end{equation}
where $U_{\rm B}$ is the total magnetic energy and $U_{\rm res}$ is the fraction of this energy resonant with particles at the given $\gamma$. This formula can be derived from the full, quasi-linear expression for the acceleration rate (as in \citealp{LAC77} or \citealp{SCH89}). Because we are most interested in the highest energies that can resonate with the turbulent spectrum, we estimate the acceleration time solely for particles with $r_{\rm g} \sim \lambda_{\rm max}$. If the turbulent spectrum decays with wavelength as described by the Kolmogorov or Kraichnan models, $\tau_{\rm res}$ increases slowly with particle energy, so $\tau_{\rm res}$ for the highest energies that `fit' is also a good characterisation of acceleration of lower-energy particles.

Pertinent to $\chem \iso{7}{}Li$ nuclei of energy 55\,EeV, $B=0.9$\,$\mu$G and $\upsilon_{\rm A} = 0.081c$, from Eq.\,\ref{eq:resonant} and disregarding $U_{\rm B}/U_{\rm res}$ follows a resonant acceleration time of $\sim10.9$\,Myr. For $\chem \iso{12}{}C$ of 55\,EeV this yields $\sim5.5$\,Myr, and for $\chem \iso{16}{}O$ of this energy we calculate $\sim4.1$\,Myr. A locally larger magnetic field, as expected in the giant lobe filaments, would lower these $\tau_{\rm res}$ results. Alternatively, considering the maximum energy of the above nuclei fitting into the maximum turbulent eddy scale in the range $30-100$\,kpc, the overall $\tau_{\rm res}$ is between $\sim14.9$ and $\sim49.7$\,Myr.

We want to compare the acceleration time for UHECR to the major loss process, namely diffusion out of the radio lobes. Because the lobes are magnetically separated from their surroundings, we consider cross-field diffusion. In a collisionless plasma this is very slow; but particle propagation in electrostatic turbulence can create much faster anomalous diffusion, also called `Bohm diffusion' (e.g. \citealp{TAY71})\footnote{The term `Bohm diffusion' is also used in a very different context, namely parallel diffusion by Alfv\'en wave scattering within magnetically connected regions. Although it is not clear that this limit is often reached in MHD turbulence (e.g. \citealp{CAS02}), Bohm diffusion is often invoked as the slowest possible diffusion in these situations.}. Following \cite{ROS93}, also \cite{BUL10}, we write the diffusion coefficient as $D_{\rm B} \simeq r_{\rm g}\,c/\xi$, where different authors find the fudge factor $\xi \sim 3 - 30$. Thus we estimate the diffusion time as
\begin{equation}
\tau_{\rm diff} \simeq \xi\,R^2/r_{\rm g}\,c \,.
\end{equation}
For numerical estimates we use $\xi\sim10$. Taking the mean of the lobe radii, $R\sim90$\,kpc, and particle gyroradii of 30 and 100\,kpc, the foregoing relation gives diffusion times of $\sim8.8$ and $\sim2.6$\,Myr respectively. 

The requirement for a relatively flat power law, as is generally assumed for UHECRs, and is also measured (\citealp{PAO10, PAO11b}), is $\tau_{\rm acc}\lesssim\tau_{\rm diff}$. For a particle gyroradius of $r_{\rm g}=30$\,kpc this is satisfied as long as the particles do not exceed $\sim44$\,EeV ($\chem \iso{7}{}Li$), $\sim59$\,EeV ($\chem \iso{9}{}Be$), $\sim89$\,EeV ($\chem \iso{12}{}C$), $\sim118$\,EeV ($\chem \iso{16}{}O$) and $\sim384$\,EeV ($\chem \iso{56}{}Fe$) in production energy, and both timescales are also amply within the estimated dynamical age (Sect.\,\ref{sect:sizeandage}) of the lobes. For a gyroradius $r_{\rm g}=100$\,kpc, the above results shift to $\sim4$\,EeV (proton), $\sim9$\,EeV ($\chem \iso{4}{}He$), $\sim13$\,EeV ($\chem \iso{7}{}Li$), $\sim18$\,EeV ($\chem \iso{9}{}Be$), $\sim27$\,EeV ($\chem \iso{12}{}C$), $\sim35$\,EeV ($\chem \iso{16}{}O$) and $\sim115$\,EeV ($\chem \iso{56}{}Fe$) production energy. 

Thus, in the context of our model, light elements at UHECR energies fit into the likely lobe turbulence driving scale well. The situation is less clear for $\chem \iso{4}{}He$ nuclei and protons; they would necessitate a driving scale of respectively $\geq33$ and $\geq66$\,kpc. The diffusion time restricts the maximum energy of $\chem \iso{4}{}He$ nuclei of $r_{\rm g}=33$\,kpc leaving the lobe to 26.8\,EeV and of protons of $r_{\rm g}=66$\,kpc to 6.7\,EeV, hence $\chem \iso{4}{}He$ and protons cannot be accelerated to the PAO band. $\chem \iso{7}{}Li$ nuclei do `fit' in our eddy size range but because of the $\tau_{\rm diff}$ constraint, they are not likely to be accelerated to the PAO band either. However, were the eddy size 24\,kpc, i.e. somewhat below our adopted range, they would make it. Finally, note that we have used an average lobe radius for the above calculations; the perspective is slightly more favourable for the southern giant lobe, but this will not alter the overall outcomes.

\subsubsection{Hybrid acceleration mechanism and seeds} \label{sect:seeds}

\cite{CHR89}, \cite{KRO04}, \cite{BEN08} and \cite{STA13} have proposed {\it in situ} magnetic reconnection as a viable mechanism for particle acceleration in galaxies' lobes. \cite{LAZ99} have shown that the magnetic reconnection is determined by the degree of field wandering induced by magnetic turbulence, and we consider the physical conditions in Cen\,A's giant lobes (see Sect.\,\ref{sect:physicsGLs}) appropriate for magnetic reconnection to exist. Based on, e.g., \cite{KOW12a} we deem it likely that magnetic reconnection and resonant acceleration co-exist. The resonant acceleration model for the giant lobes in Sect.\,\ref{sect:consequences} is subject to an `injection problem' as it comes down to the lightest nuclei\footnote{The injection energy depends on mass and charge: protons can be picked up by Alfv\'en waves at lower energies than electrons, and heavier particles have generally lower resonance thresholds than lighter ones in the (here adequate) weak scattering regime (e.g. \citealp{ZAN01}). The details of the injection energy versus mass-to-charge ratio depend sensitively on local conditions and will be investigated elsewhere.}, i.e., to accelerate those particle species beyond the energy $0.5\,m\,\upsilon_{\rm A}^2$, seeds are required. Reconnection might provide those seeds \citep{MEL83, DRU12}, as no minimum energy threshold for thermal particles is required to be accelerated by magnetic reconnection.

The presence of turbulence is expected to increase reconnection rates (\citealp{LAZ99, KOW09, KOW12b}). Assuming {\it fast} (i.e. independent of resistivity) reconnection, the reconnection velocity $\upsilon_{\rm rec}$ is taken of the order of Alfv\'en speed (\citealp{LAZ99,EYI11}), $\upsilon_{\rm rec}\!\sim\!\upsilon_{\rm A}\!\sim\!0.081c$. Under ideal circumstances, the fastest possible reconnection acceleration timescale is $\tau_{\rm rec}=r_{\rm g}/\eta\,\upsilon_{\rm A}$ \citep{GOU05}, where $\eta$ is the reconnection efficiency factor. Slower reconnection acceleration timescale occurs at particle's mean free path $\lambda_{\rm mfp}/\eta\,\upsilon_{\rm A}$ (i.e. $\lambda_{\rm mfp}>r_{\rm g}$). If we consider the former case, $\lambda_{\rm rec} = r_{\rm g}$, and adopting a gyroradius equivalent to the scale\footnote{Here we refer to the distance between the converging magnetic field lines; not to be confused with the length $L$ of a reconnection region.} of a typical reconnection region of order of $\sim1$\,pc\footnote{This is rather speculative, yet the `small' scale is justified by the requirement that a reconnection region should be $\ll$ the scale of shock structure \citep{DRU12}.}, with the proviso that $\eta\sim1$ be used (i.e., a single velocity change expected for every collision with the magnetic inhomogeneities), we obtain $\tau_{\rm rec}\sim40$\,yr, which is much faster than we have derived for the resonant acceleration in Sect.\,\ref{sect:consequences}. \cite{LAZ99} have shown that most of the energy in the reconnection is transferred into turbulent motions, and in fact, the process of the reconnection is an intrinsic part of the MHD turbulent cascade (see also \citealp{EYI11}). Given that the sizes of the reconnection regions are much smaller than the gyroradii of UHECRs in the lobes, reconnection cannot accelerate particles to energies $\ge55$\,EeV.

High-energy particles in the jets driven out by the AGN can also act as seeds, provided the particles do not lose too much energy (e.g. through adiabatic- or synchrotron losses) as they stream through the jet. We know that high-energy leptonic particle acceleration takes place in the jet from observations of high-energy synchrotron radiation. Fermi\,I acceleration at shocks and acceleration via shear at the jet boundary are themselves mediated by Alfv\'en waves, hence will have the same injection problem as turbulent Alfv\'enic acceleration. Magnetic reconnection, in the jet, or {\it in situ} in the giant lobes, may be therefore regarded as a tenable alternative. Propitious transport conditions from the jets may exist, as no photodisintegration {\it en route} of $\lesssim$\,PeV nuclei should occur at the photon number densities assumed in Cen\,A's jets and lobes (\citealp{ALL08} and references therein). 

Thus, because of the small dimensions involved, not enough energy will go into seed particles during magnetic reconnection to produce the observed UHECRs, i.e., magnetic reconnection is not useful as the main acceleration process in the lobes. However, being prompt, it is a prime candidate as the process which pre-accelerates particles to energies at which resonant acceleration becomes operational.

\section{Summary and Conclusions} \label{sect:discussion}

The main results of this paper can be summarised as follows.

1. We affirm the consistency of various estimates of the current jet power of $\sim1\times10^{43}$\,erg\,s$^{-1}$. The inclusion of special relativistic effects makes relatively little difference to the power of the jet. We have shown that in the inner jet the estimated jet power can be supplied by the observed electron population and an equipartition magnetic field alone, moving at the observed speed of $0.5c$. This does not unambiguously prove that the energetics are initially dominated by electrons and magnetic field, but it indicates that it is very likely. We have verified that the jet is not significantly over/underpressured with respect to the surrounding ISM, which allows for the Kelvin-Helmholtz instability to develop at the jet-ISM interface.

2. We have carried out rough modelling of external entrainment from hot gas using the results of \cite{LAI02a} for the FR\,I source 3C\,31. We have inferred an entrainment rate of $\sim3.0\times10^{21}$\,g\,s$^{-1}$, which is a factor of a few below the rate for 3C\,31. The fact that the jet might not be embedded in the inner lobe until $\sim3.7$\,kpc (projected) supports the credibility of our external entrainment estimates. Our internal entrainment modelling which relies on the generous assumption that the stars are not affected by the jet plasma, resulted in $\sim6.8\times10^{22}$\,g\,s$^{-1}$. This is a factor 7 below 3C\,31's internal entrainment rate, on the same assumption. The derived particle content implies imbalance between the internal lobe pressure available from relativistic leptons and magnetic field and the external pressure. To provide the requisite pressure, the material would need to be heated to $2.6\times10^{11}$\,K (southern inner lobe) and to $2.0\times10^{12}$\,K (giant lobes). 

3. The ratios between the synchrotron ages of the giant lobes of Cen\,A ($\sim30$\,Myr; \citealp{HAR09}) and our derived sound-crossing timescale ($\sim440-645$\,Myr) and buoyancy age ($\sim560$\,Myr) of the lobes are dissimilar to most age estimates through these methods in FR\,II sources, although good contraints on dynamical ages of FR\,I sources are thus far lacking. We stress that our sound-crossing timescale suffers from lack of a tight constraint on the plasma temperature surrounding the rising giant lobes, and the buoyancy age principally from an insufficient knowledge of the dominance over the lifetime of high internal pressure and/or ram presssure versus buoyancy force, and on the 3D structure of the lobes.

4. We have employed the scaling properties of the gravitational mass and X-ray emitting gas to estimate the thermal pressure and temperature of the giant lobes. Our crude modelling gives us a mean thermal pressure of $1.5\times10^{-12}$\,dyn\,cm$^{-2}$. This deduction sets the lower limit to the giant lobe temperature as $T\sim1.6\times10^8$\,K. Pressure and dynamical age considerations imply a power of the pre-existing jet (inflating the giant lobes) of $\sim5\times10^{43}$\,erg\,s$^{-1}$.

5. If the assumptions about the environment hold and if the {\it Fermi}-LAT results are correct (recall that these results only account for electrons), then we require another component in the lobes as well as electrons and magnetic field, and if this is thermal material, then the giant lobes must be dominated by thermal pressure. Even the most conservative limits on the particle density require this material to be hot, and that if it is all supplied by the entrainment it must be outstandingly hot.

6. We have presented arguments for mixed UHECR composition at the source, and suggest that thermal material in the giant lobes may well be enriched in light elements from stellar winds. Large amounts of entrainment, or in general high thermal matter content in the giant lobes, arrests UHECR production via the resonant process in the lobes; in order to accelerate hadrons to the UHE regime, the Alfv\'en speeds in the giant lobes must be mildly relativistic (a result in agreement with \citealp{HAR09} and \citealp{SUL09}). We have deduced that, to meet the pressure requirements for the lobes, the hadrons must be very hot. The same high temperatures that allow self-consistency between the entrainment calculations and the missing pressure also allow stochastic UHECR acceleration models to work. 

7. Our conclusion above is strengthened by a consistency check incorporating the turbulent properties of the giant lobes. The turbulence is sub-Alfv\'enic, yet the turbulent speed is comfortably close to the Alfv\'en speed in the lobes, hence the requirement $\upsilon_{\rm t}\sim\upsilon_{\rm A}$ is satisfied. Our computed resonant acceleration time for the lightest UHE nuclei which `fit' into the turbulent spectrum of $30-100$\,kpc in the giant lobes, and the escape time, are both comfortably within the estimated dynamical age of the lobes. In the frame of our resonance model, $\chem \iso{7}{}Li$ and heavier nuclei fit in the likely driving scale, however the diffusion time restricts the particle species accelerated to $\geq55$\,EeV to $\chem \iso{9}{}Be$ and heavier nuclei.

8. Magnetic reconnection is not expected to alleviate the hadron heating problem but it will help to lift the hadrons out of the thermal pool and, due to its promptness, will pre-accelerate particles in the giant lobes. We have pictured Cen\,A as a probable source of at least several of the UHE events detected by the large particle experiments and associate these events with light intermediate nuclei. Even considering the tangible uncertainties in some of the relevant parameter values, Cen\,A does not make the scene as a genuine producer of UHE protons.

Prospects for constraints on Cen\,A physics that may be retrieved from surveys with current and future radio, X-ray, gamma-ray and particle detection instruments include the following. Very Long Baseline Array (VLBA) circular polarisation data could provide limits on the relativistic particle population of the parsec-scale jet. Low-frequency radio polarisation observations with the Square Kilometer Array (SKA) may place stronger constraints on the thermal particle content of the jet knots and of the giant lobes. {\it XMM-Newton} observations will allow us to assess the distribution of internal energy within the lobes and may give us a more complete picture of localised particle acceleration in the large-scale lobes. {\it Suzaku} observations may constrain the plasma temperature surrounding the giant lobes, setting tighter limits on the sound-crossing timescale of the lobes. The Cherenkov Telescope Array (CTA) may provide firm limits on the TeV flux from pion decay in the giant lobes. The detection of VHE neutrinos, by ANTARES, IceCube or the PAO, with an angular resolution better than 1$^\circ$, could discriminate between neutrinos from Cen\,A's jet/core and from its giant lobes, identifying the region of their parent cosmic ray production.

\begin{acknowledgements}$\!\!\!\!$We thank G.\,Bicknell, F.\,Israel, P.\,Kronberg, I.\,Feain, O.\,Pols, R.\,Perley, E.\,K\"ording, H.\,Falcke, F.\,Owen, H.\,Jerjen, A.\,Corstanje, A.\,Pe'er, J.-P.\,Macquart, \L{}.\,Stawarz, J.\,Kuijpers, C.\,Cheung, L.\,Kaper, J.\,Sanders, L.\,Drury, H.\,Henrichs, Gopal-Krishna, A.\,Levinson, T.\,Piran, J.\,Petrovic and D.\,Harari for spirited discussions on the subject. JHC is grateful for support from the South-East Physics Network (SEPNet). AL acknowledges support from the National Science Foundation (NSF, Grant No.\,1212096).
\end{acknowledgements}

\bibliographystyle{aa} 
\bibliography{cena_he} 

\end{document}